\documentclass[12pt]{iopart}
\usepackage{epsfig}

\begin{document}

\title[Evidence for non-Gaussian tail in 3D pion emission source at
SPS]{Evidence for non-Gaussian tail in 3-Dimensional pion emission source at
SPS}

\author{P Chung$^1$ and P Danielewicz$^2$ for The NA49 Collaboration}

\address{$^1$ Dept of Chemistry, SUNY Stony Brook, 
Stony Brook, NY~11794-3400, USA}
\address{$^2$ National Superconducting Cyclotron Laboratory and
    Department of Physics and Astronomy, Michigan State University,
    East Lansing, MI~48824-1321, USA}
\ead{pchung@mail.chem.sunysb.edu}
\begin{abstract}
The NA49 experiment at CERN SPS has acquired a huge data set of Pb+Pb events over a
broad range of energy and centrality during the last several years. This high statistics data set,
coupled with a state-of-the-art analysis technique, allows for the first
model-independent extraction and energy scan of 3D emission
sources for pion pairs at SPS energies. These 3D pion emission sources provide new insights into
the nature of a long-range source previously reported by PHENIX at RHIC. The new results indicate that the pion source displays significant non-Gaussian tails in the longitudinal direction at 40 and 158~AGeV and in the
outward direction at 158~AGeV.

\end{abstract}

\pacs{25.75.-q, 25.75.Gz, 25.70.Pq}

\section{Introduction}
      A deconfined phase of nuclear matter is expected to be formed at the
 high energy densities created in relativistic heavy ion collisions \cite{qgp03}. It is
 widely believed that important signatures of such a phase
 are reflected in the space-time extent and shape of particle emission
 source functions.
   
     Recently, 1-Dimensional source imaging techniques \cite{brown97,brown98} have revealed a
 non-trivial long range structure in the two-pion emission source at RHIC
\cite{chung05,chung06}. The origins of this structure are still unclear.
 The presence/absence of such a structure in the pion emission sources in
 heavy ion collisions at intermediate SPS energies could yield important
 information which could help resolve the structure's origins.
      The NA49 Collaboration has carried out Pb+Pb collisions over a
 wide range of bombarding energies at the CERN SPS during the last decade \cite{alt05}. Such a rich data set provides a unique opportunity to search for this long range structure at the SPS and study its evolution with beam energy with a
view to unraveling its nature.
  
In this paper, the 3-Dimensional emission source images for pions
 produced in central Pb+Pb collisions over the incident energy range 40 and 158~AGeV are presented. The results are discussed in the context
 of a Gaussian shape assumption.
 
\section{Experimental Setup and Data Analysis}
  
The data presented here were taken by the NA49 Collaboration during the years 1996-2002. The incident beams of 40 and 158~AGeV were provided by the CERN SPS accelerator. The NA49 Large Acceptance Hadron Detector \cite{afa99}
achieves large acceptance precision tracking ($\delta p/p^2 \approx (0.3-7).10^{-4} (GeV/c)^-1$) and particle identification using time projection chambers.
Charged particles are detected by tracks left in the TPC and identified by the energy deposited in the TPC gas. Mid-rapidity particle identification is further enhanced by a time-of-flight wall (resolution 60ps). Event centrality is determined by a forward calorimeter which measures the energy of spectator matter.
                                                                                
3D correlation functions , C($\mathbf{q}$), were obtained as the ratio of pair to uncorrelated reference distributions in relative momentum $\mathbf{q}$ for $\pi^-\pi^-$ pairs. Here, $\mathbf{q}=\frac{(\mathbf{p_1}-\mathbf{p_2})}{2}$ is half of the relative
momentum between the two particles in the Pair Center-of-Mass System (PCMS). The pair
distribution was obtained using pairs of particles from the same event and the uncorrelated distribution
was obtained by pairing particles from different events. Track
merging and splitting effects were removed by appropriate cuts on both the pair
and uncorrelated distributions. Momentum resolution effects were negligible.
  
In the cartesian harmonic decomposition technique \cite{daniel05,chung05}, the 3D correlation function is expressed as
\begin{equation}
C(\mathbf{q}) - 1 = R(\mathbf{q}) = \sum_l \sum_{\alpha_1 \ldots \alpha_l}
   R^l_{\alpha_1 \ldots \alpha_l}(q) \,A^l_{\alpha_1 \ldots \alpha_l} (\Omega_\mathbf{q})
\end{equation}
where $l=0,1,2,\ldots$, $\alpha_i=x, y \mbox{ or } z$, $A^l_{\alpha_1 \ldots \alpha_l}(\Omega_\mathbf{q})$
are cartesian harmonic basis elements ($\Omega_\mathbf{q}$ is solid angle in $\mathbf{q}$ space) and $R^l_{\alpha_1 \ldots \alpha_l}(q)$ are cartesian correlation moments given by
\begin{equation}
 R^l_{\alpha_1 \ldots \alpha_l}(q) = \frac{(2l+1)!!}{l!}
 \int \frac{d \Omega_\mathbf{q}}{4\pi} A^l_{\alpha_1 \ldots \alpha_l} (\Omega_\mathbf{q}) \, R(\mathbf{q})
 \label{eqn2}
\end{equation}
The coordinate axes are oriented so that z is parallel to the beam (long) direction, x points
in the direction of the total momentum of the pair in the Locally Co-Moving System (LCMS) (out) and y is perpendicular to the other two axes (side).
  
The correlation moments, for each order $l$, are calculated from the measured 3D correlation function using equation (\ref{eqn2}). Each independent correlation moment is then imaged using the 1D Source Imaging code of Brown and Danielewicz \cite{brown97,brown98} to obtain the corresponding source moment for each order $l$. Bose-Einstein symmetrisation and Coulomb interaction (the sources of the observed correlations) are contained in the source imaging code.
Thereafter, the total source function is constructed by combining the source
moments for each $l$ as in equation (\ref{eqn3})
  
\begin{equation}
 S(\mathbf{r}) = \sum_l \sum_{\alpha_1 \ldots \alpha_l}
   S^l_{\alpha_1 \ldots \alpha_l}(r) \,A^l_{\alpha_1 \ldots \alpha_l} (\Omega_\mathbf{r})
\label{eqn3}
\end{equation}

\section{Results}
  
\begin{figure}
\includegraphics[width=0.46\linewidth]{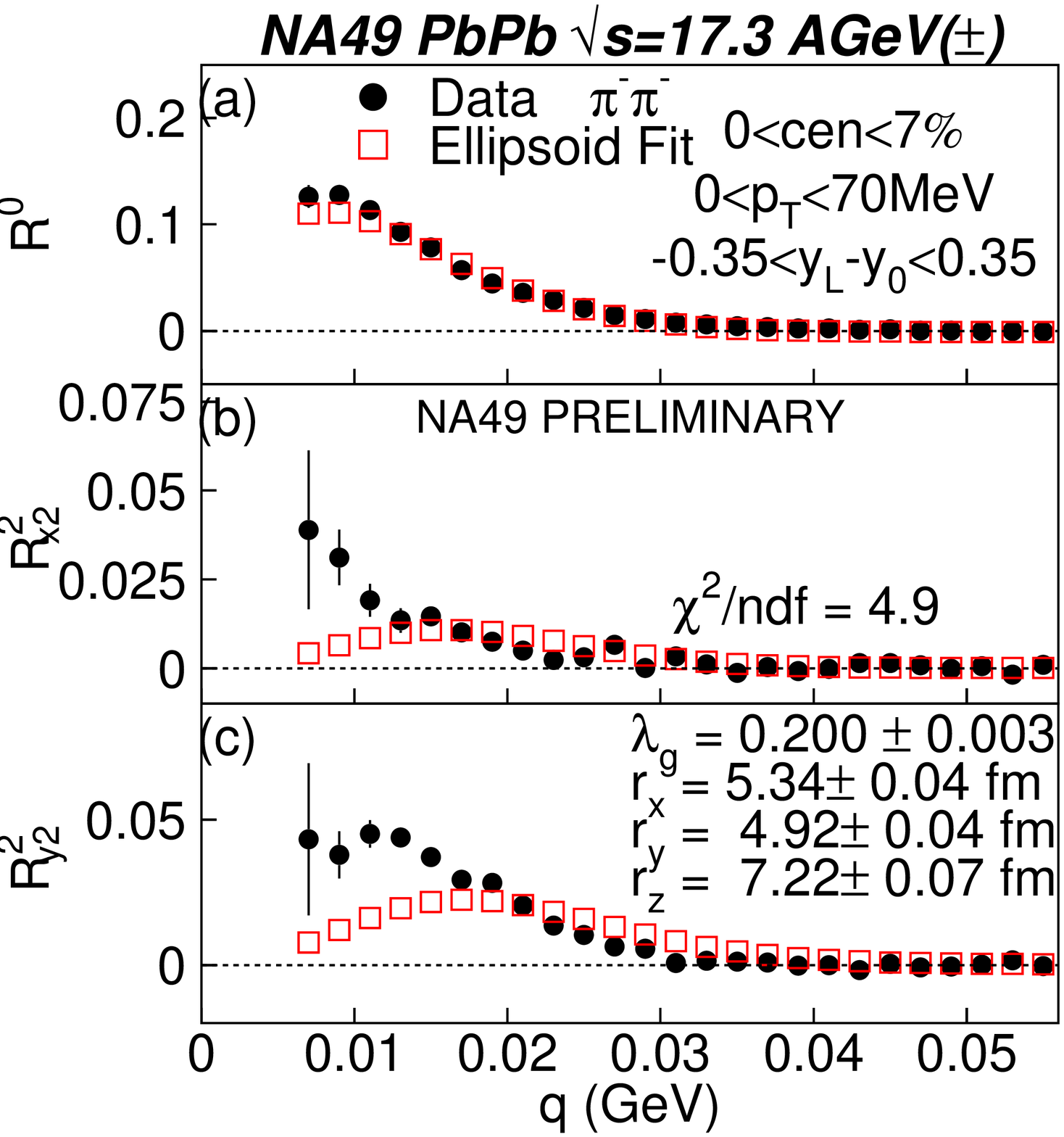}
\includegraphics[width=0.46\linewidth]{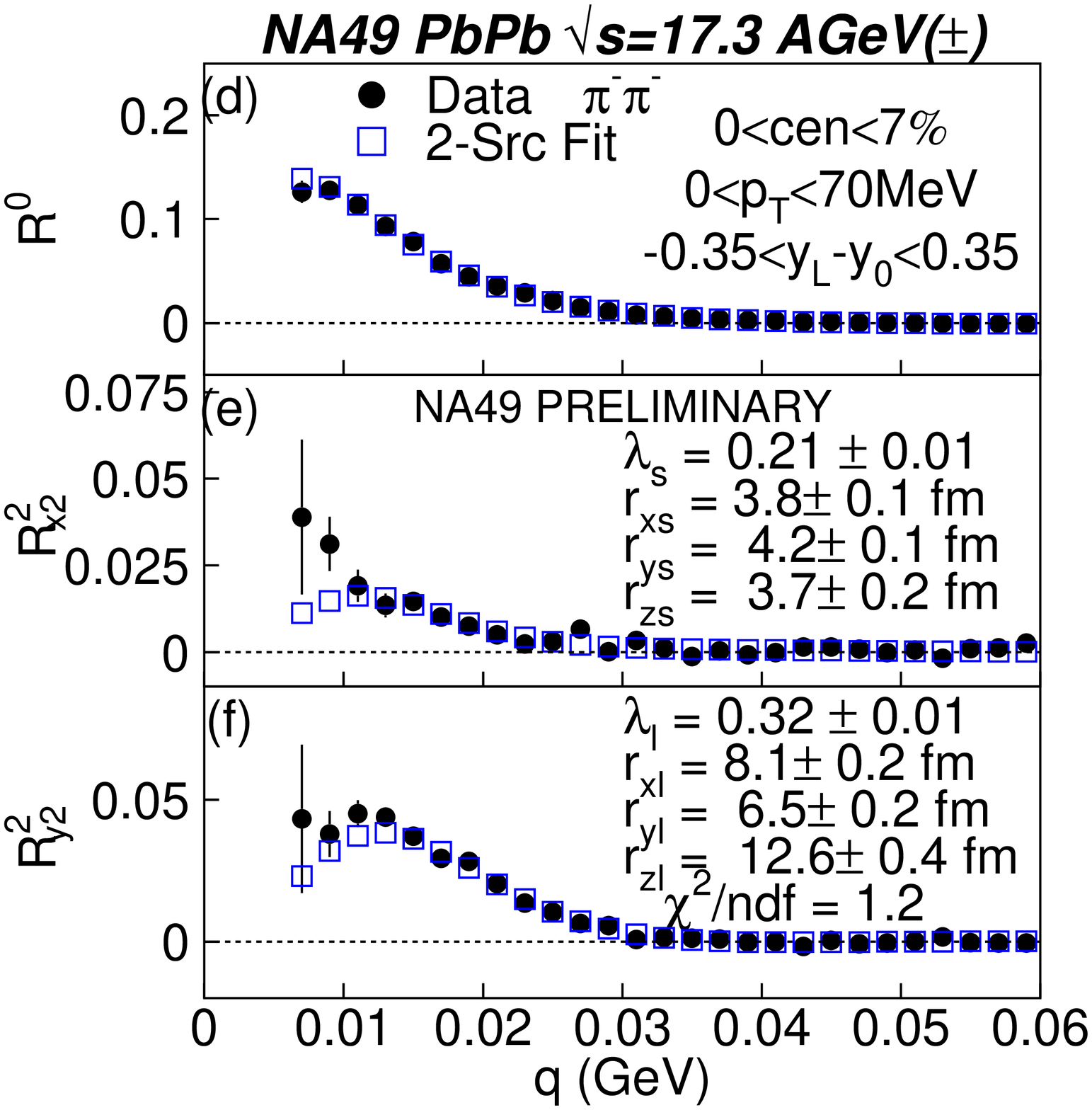}
\vskip -1.cm
\caption{\small{$l = 0$ and $l = 2$ moments for mid-rapidity low $p_T$ $\pi^-\pi^-$ pairs from 158~AGeV central Pb+Pb collisions as a function q. Data are shown as solid circles while the
squares represent the result of a simultaneous fit of the moments with an ellipsoid shape ((a)-(c)) and a 2-source model ((d)-(f)).}}
\label{40_160gev_fit}
  
\end{figure}

Figure \ref{40_160gev_fit} shows the $l = 0$ ($R^0$) and $l = 2$ ($R^2_{x2}$ and $R^2_{y2}$) moments for mid-rapidity ($|y_L-y_0|<0.35$, where $y_L$ is particle laboratory rapidity and $y_0$ is CM rapidity), low $p_T$
($0<p_T<70MeV/c$) $\pi^-\pi^-$ pairs from 158~AGeV central (7$\%$)
 Pb+Pb collisions, as a function of the relative momentum q in the PCMS frame. Here, $R^2_{x2}$ is shorthand for $R^2_{xx}$ etc. Moments for other l's are either zero or negligible. The Lorentz transformation of $\mathbf{q}$ from the laboratory frame to the PCMS is done by transforming to the LCMS along the beam axis and then transforming to the PCMS along the pair transverse momentum.
 
The data are represented in Fig. \ref{40_160gev_fit} by solid circles, while squares represent the results of fits to the independent moments with model sources, an ellipsoid shape (3D Gaussian) in panels (a)-(c) and a 2-source shape in panels (d)-(f). The last function arises by assuming a linear combination of two Gaussians for the single particle distribution. The best-fit
 parameters for the two shapes are listed in the lower panels. The 2-source shape yields a much better representation of the data. 

\begin{figure}
\includegraphics[width=0.46\linewidth]{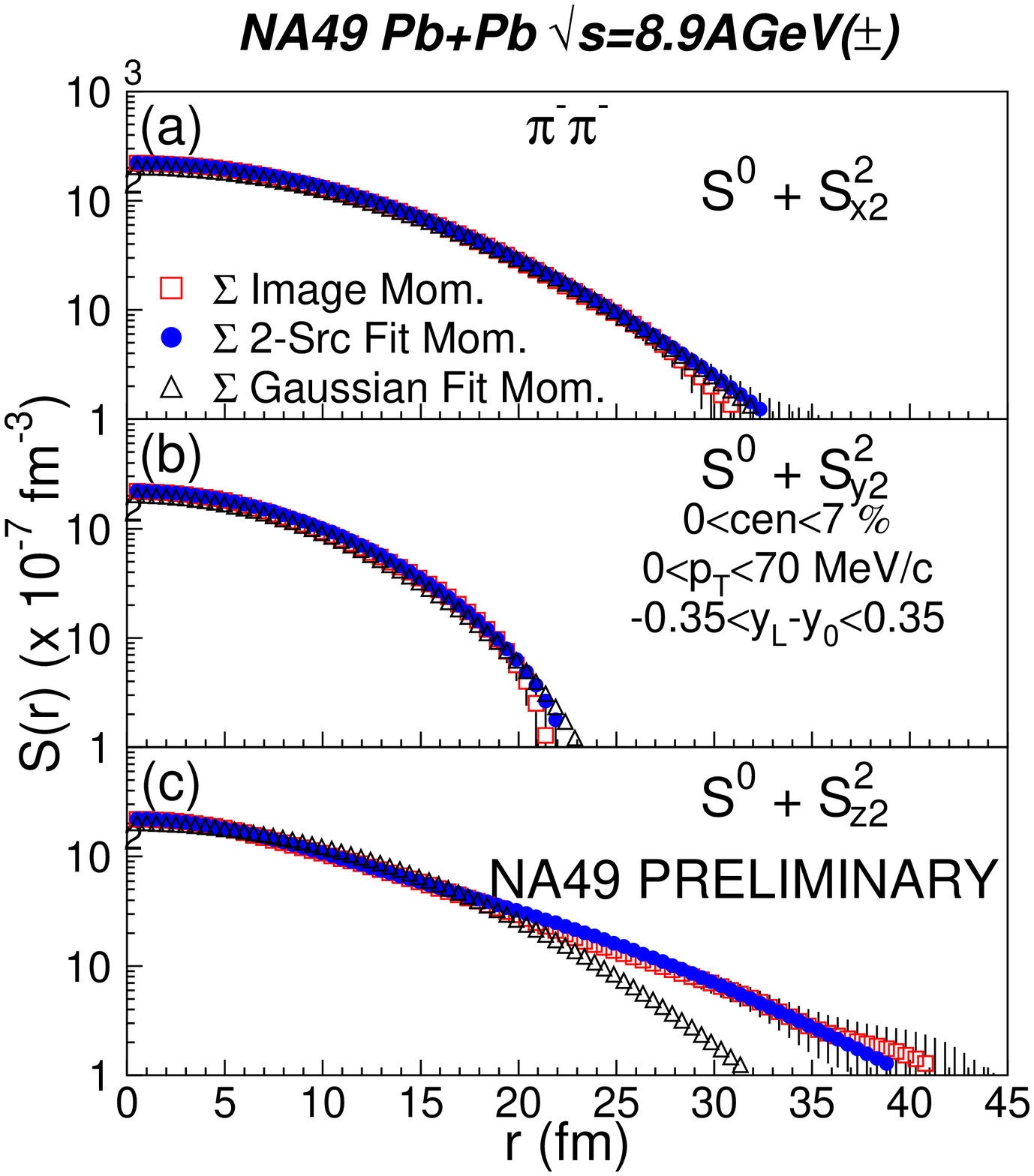}
\includegraphics[width=0.46\linewidth]{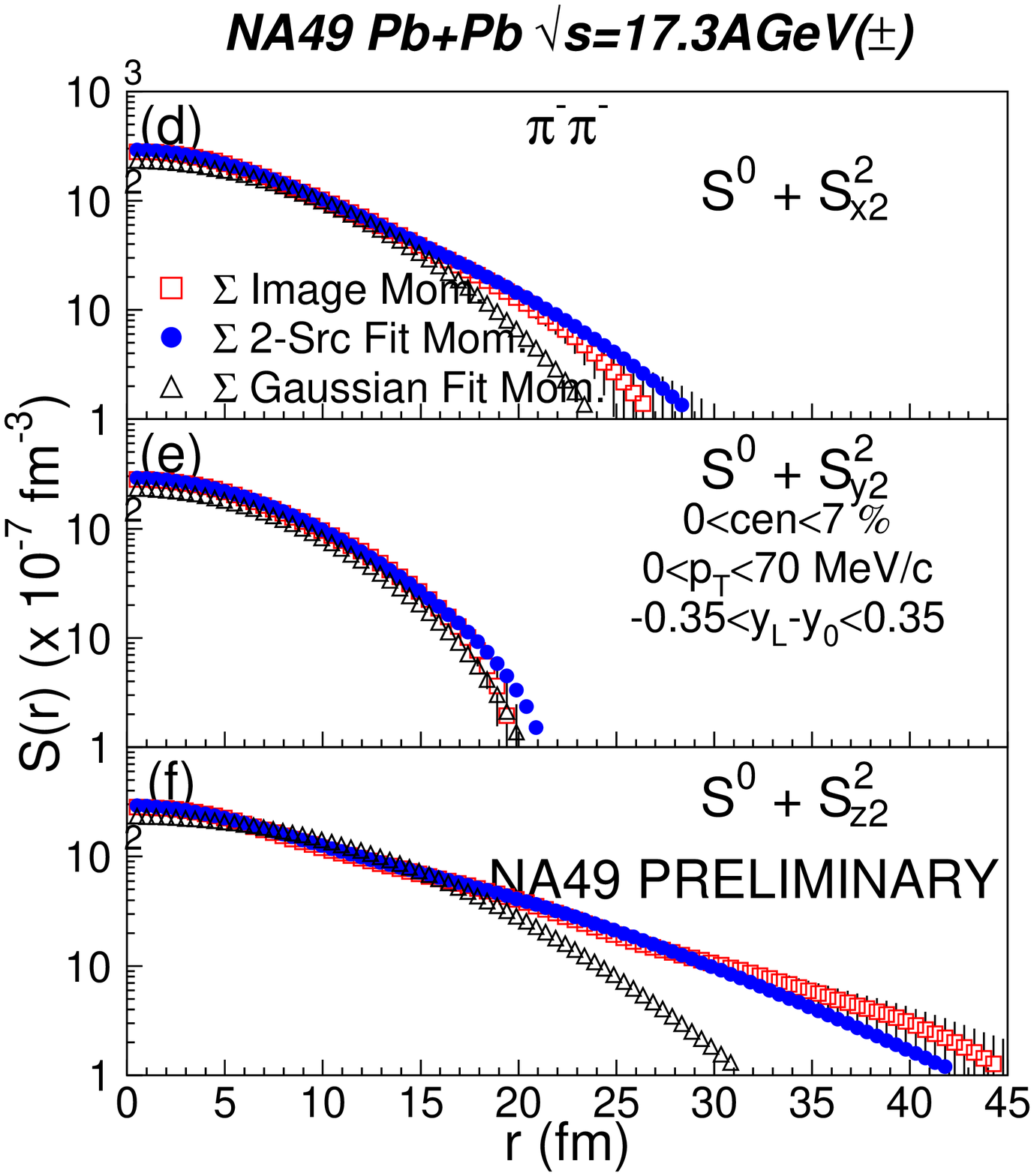}
\vskip -1.cm
\caption{\small{Source function in x (out), top panels, y (side), middle panels, and z (long)
directions for low $p_T$ mid-rapidity $\pi^-\pi^-$ pairs from central
Pb+Pb events at $\sqrt s = 8.9 $AGeV (incident beam energy of 40~AGeV)
and $\sqrt s = 17.3 $AGeV (incident beam energy of 158~AGeV). The imaged,
Gaussian and 2-source functions are represented, respectively, by squares,
circles and triangles.}}
\label{40_160gev_src}
\end{figure}

Given that the only significant imaged moments are found for $l=0$ and 2 multipolarities, the net imaged source function in the x, y and z directions is
simply the sum of the 1D source $S^0$ and the corresponding $l=2$
 moment $S^2_{ii}$ where i= x, y or z. In panels (a)-(c), Figure \ref{40_160gev_src} compares the net imaged 
source (squares) for mid-rapidity, low $p_T$ pion pairs from central Pb+Pb collisions at 40~AGeV (i.e $\sqrt s=6.4$AGeV) to the best-fit parametrized functions, Gaussian (triangles) and 2-source model (circles).
 The source image and the 2-source model agree very well in all 3 directions and both disagree with the Gaussian fit in the z direction.

 At 158~AGeV (i.e $\sqrt s=17.3$AGeV), represented in panels (d)-(f) of Fig.
\ref{40_160gev_src}, the findings are similar, but in addition to the z direction, the source function starts to develop a non-Gaussian tail in the x direction.

\section{Discussions}

The ratio of the RMS radii of the source functions in the x and y directions is 1.3$\pm$0.1 at 40~AGeV and 1.2$\pm$0.1 at 158~AGeV. This deviation from unity, evident visually, points to a finite pion emission time. 

Moreover, the RMS pair separation in the z direction, from figure \ref{40_160gev_src}(c) and (f), is 11fm at 40~AGeV and 12fm at 158~AGeV. These dimensions are much smaller than the Lorentz-contracted nuclear diameters of 3fm at 40~AGeV and 1.5fm at 158~AGeV and in fact give the RMS pair separation due to the longidutional spread of 
nuclear matter created by the passage of the two nuclei. Since the latter
 are moving with almost the speed of light, one can infer the lower bound formation time of the created nuclear matter to be 8fm/c at 40~AGeV and 10fm/c at 158~AGeV.


\section*{References}
\begin{thebibliography}{10}

\bibitem{qgp03} QM2002, Nucl. Phys. {\bf A 715}, 1c (2003)
\bibitem{chung05} P.~Chung et al, Nucl. Phys. {\bf A 749}, 275c (2005)
\bibitem{chung06} S.S.~Adler et al.(PHENIX Collaboration), nucl-ex/0605032
\bibitem{brown97} D.A.~Brown and P.~Danielewicz, Phys. Lett. {\bf B 398}, 252 (1997)
\bibitem{brown98} D.A.~Brown and P.~Danielewicz, Phys. Rev. {\bf C 57}, 2474 (1998)
\bibitem{alt05} C.~Alt et al.(NA49), CERN-SPSC-2005-041,
CERN-SPSC-P-264-ADD-12, Nov 2005.
\bibitem{afa99} S.~Afanasiev et al., Nucl. Instrum. Meth. {\bf A430} 210 (1999).
\bibitem{daniel05} P.~Danielewicz and S.~Pratt, Phys. Lett. {\bf B 618} 60 (2005).
\end{thebibliography}
\end{document}